# SeqPoint: Identifying Representative Iterations of Sequence-based Neural Networks

Suchita Pati[*1], Shaizeen Aga[2], Matthew D. Sinclair[1, 2], and Nuwan Jayasena[2]
[1]University of Wisconsin-Madison
{spati, sinclair}@cs.wisc.edu
[2]Advanced Micro Devices, Inc.
{shaizeen.aga, nuwan.jayasena}@amd.com

***Abstract*—The ubiquity of deep neural networks (DNNs) continues to rise, making them a crucial application class for hardware optimizations. However, detailed profiling and characterization of DNN training remains difficult as these applications often run for hours to days on real hardware. Prior works have exploited the iterative nature of DNNs to profile a few training iterations to represent the entire training run. While such a strategy is sound for networks like convolutional neural networks (CNNs), where the nature of the computation is largely input independent, we observe in this work that this approach is sub-optimal for sequence-based neural networks (SQNNs) such as recurrent neural networks (RNNs). The amount and nature of computations in SQNNs can vary for each input, resulting in heterogeneity across iterations. Thus, arbitrarily selecting a few iterations is insufficient to accurately summarize the behavior of the entire training run.**

**To tackle this challenge, we carefully study the factors that impact SQNN training iterations and identify *input sequence length* as the key determining factor for variations across iterations. We then use this observation to characterize all iterations of an SQNN training run (requiring no profiling or simulation of the application) and select representative iterations, which we term *SeqPoints*. We analyze two state- of-the-art SQNNs, DeepSpeech2 and Google's Neural Machine Translation (GNMT), and show that SeqPoints can represent their entire training runs accurately, resulting in geomean errors of only 0.11% and 0.53%, respectively, when projecting overall runtime and 0.13% and 1.50% when projecting speedups due to architectural changes. This high accuracy is achieved while reducing the time needed for profiling by 345x and 214x for the two networks compared to full training runs. As a result, SeqPoint can enable analysis of SQNN training runs in mere minutes instead of hours or days.**

## I. Introduction

Deep neural networks (DNNs) are becoming increasingly popular and are used in a wide variety of application domains. As a result, DNNs represent an important driver of future hardware and software stacks. A key tool in the repertoire of a computer science researcher in catering to such trends is profiling and characterizing of program behavior as it is often a crucial step in the path to identifying software optimizations and designing hardware enhancements.

[*]This work was done while the author was an intern with AMD Research.

However, profiling DNN training is challenging given training on real-world datasets takes hours to days on real hardware. Even after discounting the often-significant overheads of profiling tools, this implies that a simple change in underlying hardware or software configuration can necessitate expensive re-profiling. Such re-profiling may often not be practical, especially given the rapid pace of evolution in DNN-oriented hardware and software platforms, leaving application designers with stale or inaccurate information. In addition, the complex software stack DNNs employ make it challenging to port end-to-end networks on architectural simulators in order to study and characterize their behavior.

To address this challenge, prior work [1] harnessed the iterative nature of DNN training to profile a few iterations after an initialization or warm-up phase, and considered these iterations representative of the overall training run. While this approach works well for DNNs such as convolutional neural networks (CNNs) where the amount and nature of computations are independent of the inputs, it is inadequate for the increasingly important class of sequence-based neural networks (SQNNs), such as recurrent neural networks (RNNs). Unlike CNNs, the amount and nature of computations in SQNNs vary with the inputs, resulting in heterogeneous iterations during the course of a training run.

One possible approach to address this heterogeneity is to create a representative training run comprised of all variants of iterations observed for a dataset and use it to profile/characterize the corresponding training run. However, as we will show in this work, due to the input dependent nature of SQNNs and the wide variety of inputs in realistic datasets, a large number of iterations are required to create such a representative set rendering this naive solution impractical.

We take a different approach, focused on exploiting the underlying features of the algorithms to identify a small subset of the training iterations that can accurately summarize the overall DNN training run. To this end, we characterize the factors that affect the execution profile of training iterations for two popular SQNNs: DeepSpeech2 (DS2) [2] and Google's Neural Machine Translation (GNMT) [3]. Our characterization results for DS2 and GNMT show that the *input sequence-length* of an iteration (e.g., number of words in input sentences to a language translation model) is the key factor that leads to variations in an iteration's execution profile. As such, exercising a curated set of sequence lengths in the training dataset can help us create a representative training run.

In order to select such a set of sequence lengths, we make the key observation that inputs of similar sequence lengths have similar execution profiles. In tandem with this observation, we extend ideas from the well-known SimPoint [4] approach to cluster sequence lengths together. Then, we pick a representative sequence length from each cluster, which we call ***SeqPoint***. Similar to SimPoints, we assign weights to SeqPoints and use the weighted sum/average of SeqPoints to project the behavior of the overall training run.

We compare SeqPoint-based projections to measurements of full training runs and show that our proposal can accurately summarize the entire training while significantly reducing the number of iterations that need to be profiled. Moreover, SeqPoint can be used as a stepping stone to further distill representative portions and simulate complex SQNNs on architecture simulators paving the path to even more detailed study of complex end-to-end networks.

The key contributions of this work are as follows:

- We show that techniques developed to identify representative training execution for CNNs are not well suited to SQNNs. SQNNs, such as RNNs, manifest heterogeneous iterations which make it challenging to pick representative iterations within the training phase.
- We study the factors that affect the execution profile of iterations in two popular SQNNs, DeepSpeech2 and GNMT, reference networks from the MLPerf benchmark suite [5].
- We identify the wide variability of *input sequence length* as the key factor leading to variations in the execution profile of SQNN training iterations.
- As input sequence-length space can be large for a SQNN training run, we develop a methodology to identify a small number of representative sequence lengths, termed *SeqPoints*, which represent the whole training run.
- We show that SeqPoints exhibit much greater accuracy in projecting overall training behavior compared to prior work's approach of selecting arbitrary iterations, and can greatly reduce profiling overheads from a matter of hours or days to mere minutes.

## II. Background

In recent years, there has been a large amount of work optimizing CPUs, GPUs, and accelerators for machine intelligence (MI) workloads. Much of this work has focused heavily on optimizing for CNNs [6]–[13]. A given CNN instance typically consumes fixed size inputs (e.g., images scaled to a specific resolution). As a result, the amount of computation performed for each input is the same, for example, with the same number of same-sized matrix or convolutional operations. This regularity across inputs relaxes to some degree with optimizations such as exploiting sparsity but, even then, the number of operations at a macro level (e.g., the number of matrix or convolution operations) remain relatively unchanged.

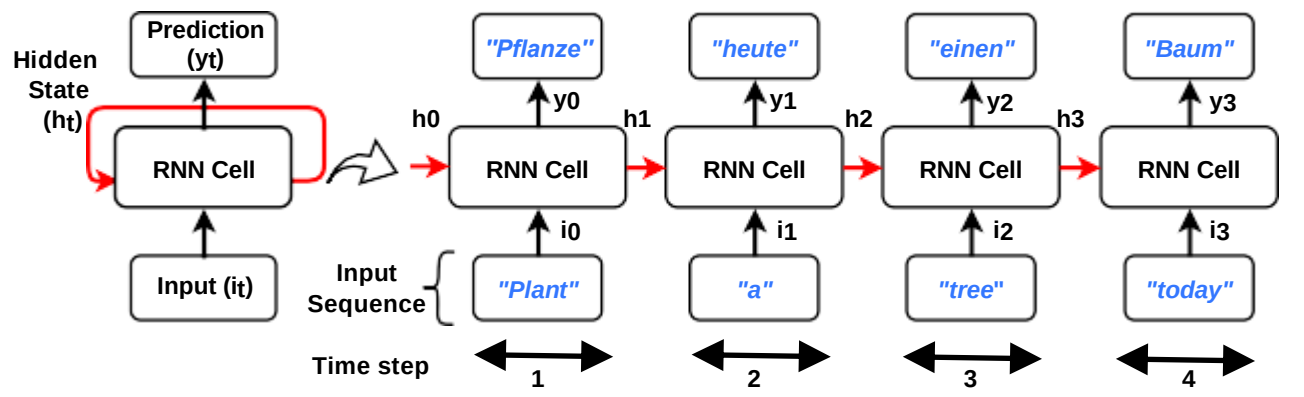

Figure 1: Left: A single layer of an RNN. Right: Unrolled RNN processing an input sequence.

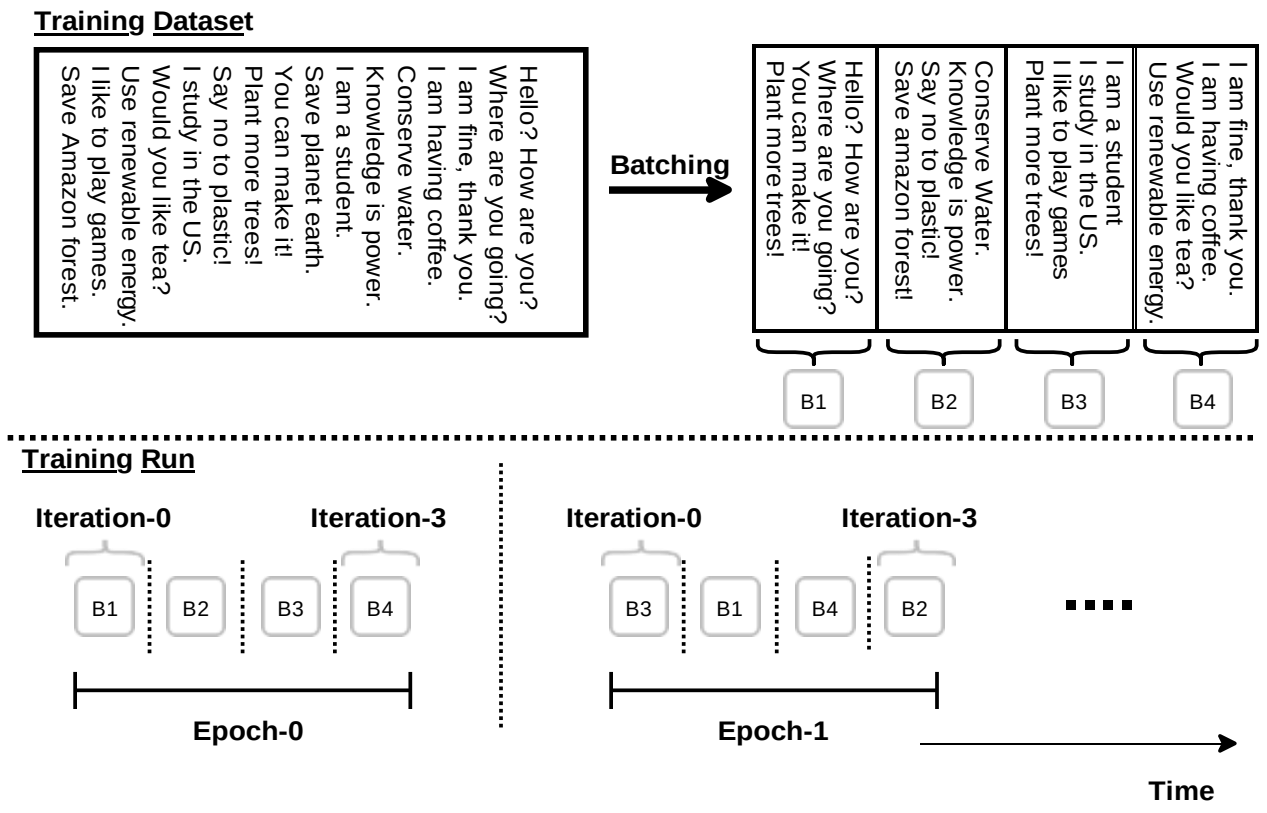

Figure 2: Training phase of an RNN.

SQNNs, such as RNNs, are another important class of networks that form the basis for many MI workloads [14]–[17], including most natural language tasks. Unlike CNNs, SQNNs perform varying amounts of computation based on the length of each input (also referred to as the sequence length). For example, in a word-granularity text-based application, the amount of computation varies with the number of words in a sentence. This variability manifests as the *cells* of the network being *unrolled* by as many *steps* as the sequence length. Further, RNNs retain state through the sequence of operations performed for each input, with each cell feeding back internally retained state from the previous step(s), in addition to consuming the next segment of the input. A simplified RNN cell is illustrated in the left side of Fig. 1, with the right side showing the unrolled view for an input with four words.

Most DNNs have large numbers of tunable parameters (or *weights*) that are learned using large amounts of data during a *training* phase. Once trained, the network can be deployed to operate on new inputs, which is referred to as *inference*. While there are multiple approaches to training a network, we focus on supervised learning as it is one of the most widely used and mature approaches, making it an important workload for architectural analysis. In the training phase using supervised learning, an input (e.g., an image) is fed into the network and is propagated forward through a collection of layers that form the network until an output is generated. Each layer takes the output of one or more previous layers, computes on it and feeds its result to the next layer. At the end of this forward-pass, the generated output is compared to the known correct output to compute an error. The error generated in the forward-pass is propagated backwards through the layers of the network during a backward-pass of the network, modifying the tunable parameters of each layer to minimize the error.

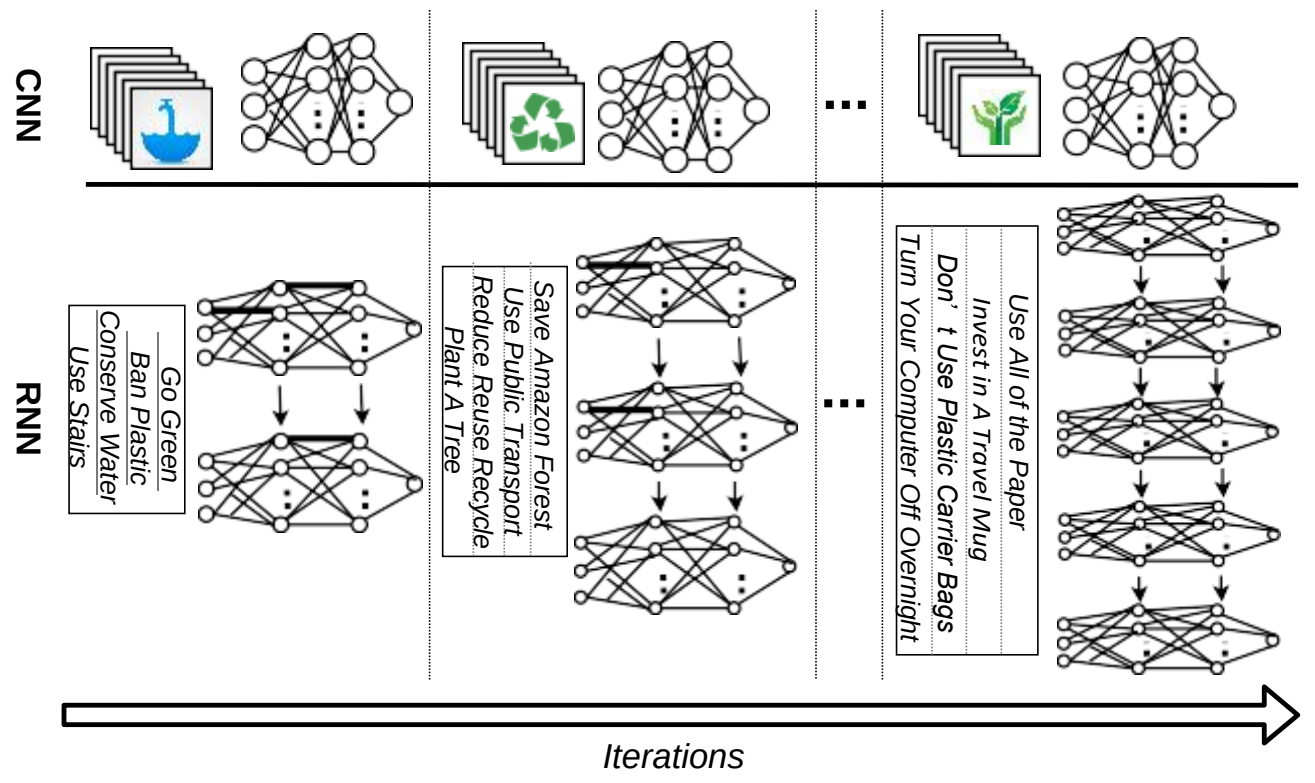

Figure 3: Comparing iterations of CNNs and SQNNs.

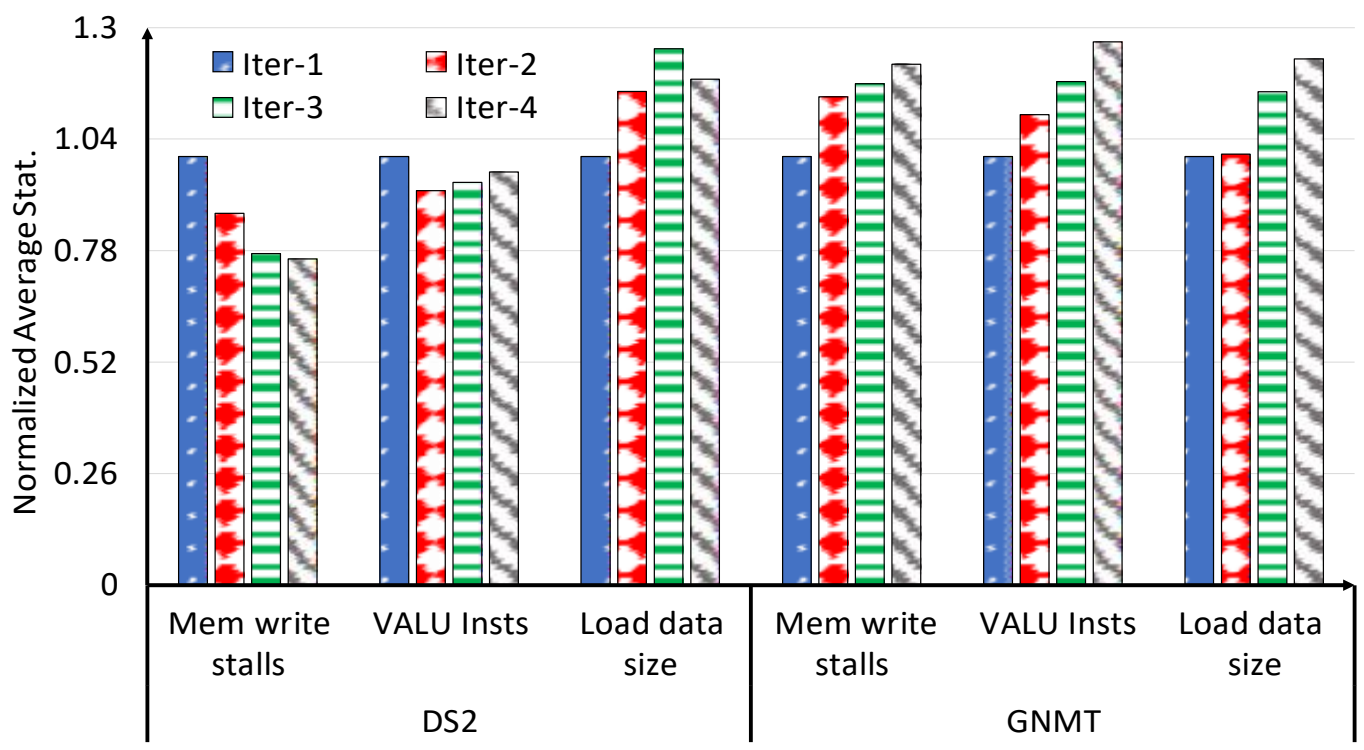

Figure 4: Architectural statistics for four representative training iterations.

To improve hardware utilization (particularly on parallel platforms such as GPUs) and to improve the stability of convergence, the training phase is often performed in groups of inputs known as *minibatches* or, simply, *batches*.[1] The number of inputs in a batch is referred to as the batch size. The upper part of Fig. 2 illustrates an example of forming batches of size four for a text-based training set. In batch-based training, all inputs of a batch perform the forward traversal using the same set of weights, and then the backward pass is performed for all inputs of the batch, computing the corresponding errors and updating the weights. The forward and backward traversal of a single batch through the network is referred to as an *iteration*. A set of iterations making up a single pass through the entire dataset is referred to as an *epoch*, as illustrated by the lower part of Fig. 2. Training of a network typically consists of multiple epochs (i.e., multiple passes over the entire training set) until a convergence criterion is met.

## III. Motivation

Profiling and characterization of application behavior forms the groundwork that guides optimizations at various levels of the hardware-software stack from architecture to system design to compilers and libraries. Given the importance of understanding program behavior, there exist a plethora of tools and techniques that work at various levels of the system stack and provide the necessary insights which help researchers and developers design the next-generation of architectural, system level, and software optimizations.

However, for complex workloads such as DNNs, application characterization is difficult due to large datasets and long runtimes. Specifically, DNN training can often take several hours to days to run on real hardware, making detailed profiling impractical even when discounting the overheads of profiling tools. Furthermore, given the complex software stack such networks are based on, it is often challenging to reproduce their execution environment and run realistic workloads with real-world datasets on architectural simulators.

A workaround for this challenge is identifying representative portions of program execution and using their characteristics to guide whole program optimization [4], [18]. While selecting representative portions is difficult because the behavior of programs change over time, past work [1] has exploited the iterative nature of DNNs (Section II) to pick a few iterations of the training phase as representative of the entire training run.

We first make the key observation that while the above strategy is sound for CNNs, where the characteristics of the computation is largely the same across different inputs, it is sub-optimal for SQNNs where the amount and nature of computation can vary with each input. Fig. 3 depicts this fundamental difference between CNNs and SQNNs such as RNNs. As discussed in Section II, the training phase of DNNs can be viewed as a collection of iterations, each with its own input batch of data. While the input batch does not affect the computation performed for CNN training, the unroll factor of RNNs (Section II) is dictated by the input batch. This leads to heterogeneous iterations for SQNNs with differing computations unlike the homogeneous iterations of CNNs training.

The heterogeneity of iterations in SQNNs is manifested in their architectural behavior as depicted in Fig. 4. In the

[1] In some contexts, particularly with regard to gradient descent, the term batch may be used to refer to the entire dataset. However, as is more typical when discussing neural networks in general, we will use batch and minibatch interchangeably.

figure, we compare a few hardware performance counter metrics (averaged across all operations) for four representative iterations of DS2 and GNMT (methodology discussed in Section VI). Specifically, we show the read memory traffic, memory write stall behavior, and vector ALU instructions. These statistics differ by about 24%, 25%, and 27%, respectively, across iterations for the networks studied. Thus, generalizing the entire training run based on a few arbitrarily-selected training iterations will likely be inaccurate since the selected iterations either may not represent iterations that have a major impact on the overall training run or will have different behavior from other iterations.

Since iterations in SQNNs are heterogeneous, to truly characterize the training phase of an SQNN, a potential strategy is to profile a single training epoch instead of the entire training run. This can suffice as epochs are largely homogeneous and encompass all possible iterations as discussed in Section II. However, even a single training epoch for complex DNNs such as DS2 and GNMT can possibly run hours to days on real-world datasets making profiling an entire epoch impractical.

Finally, unlike prior works [15], [19] which primarily focus on specific layers within SQNNs, we focus on profiling and characterizing end-to-end network training. An end-to-end SQNN, such as DS2, often comprises several heterogeneous layers in a specific configuration (e.g., convolution, batch-normalization, and GRU). Thus, characterizing individual layers often misses out on interactions between such heterogeneous layers.

In summary, existing mechanisms to profile and characterize SQNN training phase remain either inadequate or impractical. We aim to tackle this challenge by identifying representative iterations whose characteristics can accurately summarize the behavior of the entire training phase for SQNNs.

## IV. Characterizing Iteration Execution Profile

The discussion in Section III illustrated that the heterogeneity of training iterations in SQNNs makes it difficult to select arbitrary training iterations and consider their behavior representative of the training run. Thus, we must carefully select iterations that are representative of the behavior of the entire training run. Accordingly, we analyze the applications to deduce key factors that decide an iteration's execution profile and use this to select representative iterations for the training phase.

### A. Execution Profile

The execution profile of an iteration is directly related to the computations it executes. As training of SQNNs is typically executed on accelerators such as GPUs [1], in this work, we discuss execution profile in the context of GPU computations. Computations on a GPU are typically invoked as 'kernels' (analogous to functions in CPU parlance). As such, the execution profile of an iteration comprises the distribution of invoked kernels and their runtimes. Considering both of these helps us define the execution profile of an iteration.

Table I: Dimensions for the same GEMM operation across two iterations.

| | | M | K | N | |
|---|---|---|---|---|---|
| | | | | sl-1 | sl-2 |
| GNMT | GEMM-a | 36549 | 1024 | 6016 | 576 |
| | GEMM-b | 1024 | 36549 | 6016 | 576 |
| DS2 | GEMM-a | 29 | 1600 | 25728 | 3776 |
| | GEMM-b | 1600 | 29 | 25728 | 3776 |

### B. Factors Determining Execution Profile

*1) Sequence Length:* As discussed in Section II, the computations in an SQNN iteration largely decide its execution profile (i.e., the kernels and their runtimes). These computations in turn are determined by network dimensions (e.g., number of layers, hidden state size) and inputs to an iteration. As such, an iteration's execution profile is largely dictated by the network dimensions and inputs to the iteration.

Throughout a training run, the network dimensions stay constant. However, inputs vary per iteration, and are dictated by batch size and, for SQNNs, the length of the input sequences. Although the sequence length (SL) may vary for each element of a batch, most SQNNs will pick a single SL for the entire batch (usually the longest SL in the batch) and pad the remaining elements. Accordingly, while batch size is kept constant throughout the training run, the input SL varies per batch based on the specific inputs.

Input SL can affect the execution profile of an iteration in many different ways as we discuss next. Note that, while the following observations are pervasive across iterations, we only show data corresponding to a few kernels and a few iterations for brevity.

First, some layers (e.g., attention, fully-connected classifier) in a heterogeneous SQNN process the entire input sequence causing the inputs to such layers (and their operations) to differ across iterations with different SLs. Table I depicts the input matrix sizes (M, N, K) for two such GEMM operations (GEMM-a, GEMM-b) across two iterations. The matrix dimensions differ and consequently, their runtime and contribution to the overall execution profile differ. The rest of the layers (e.g., GRUs, LSTMs) usually process one symbol of the sequence at a time, hence have fixed-size inputs across iterations.

Second, due to the variation in input sizes to operations, different kernels (optimized for different input sizes) may get invoked across different iterations. Fig. 5 illustrates the proportion of unique kernels invoked across two iterations of GNMT and DS2. Each bar represents the breakdown of unique kernels for a pair of iteration, with *common* referring to unique kernels invoked for both the iterations and *only-in-x* referring to unique kernels invoked in either of the iterations. While there can be several unique kernels common to both the iterations, there are up to 20% of unique kernels which are present only in one of the iterations. Moreover, this data does not account for the dynamic invocations or the input sizes (thus runtime) of these kernels, which can also vary across iterations.

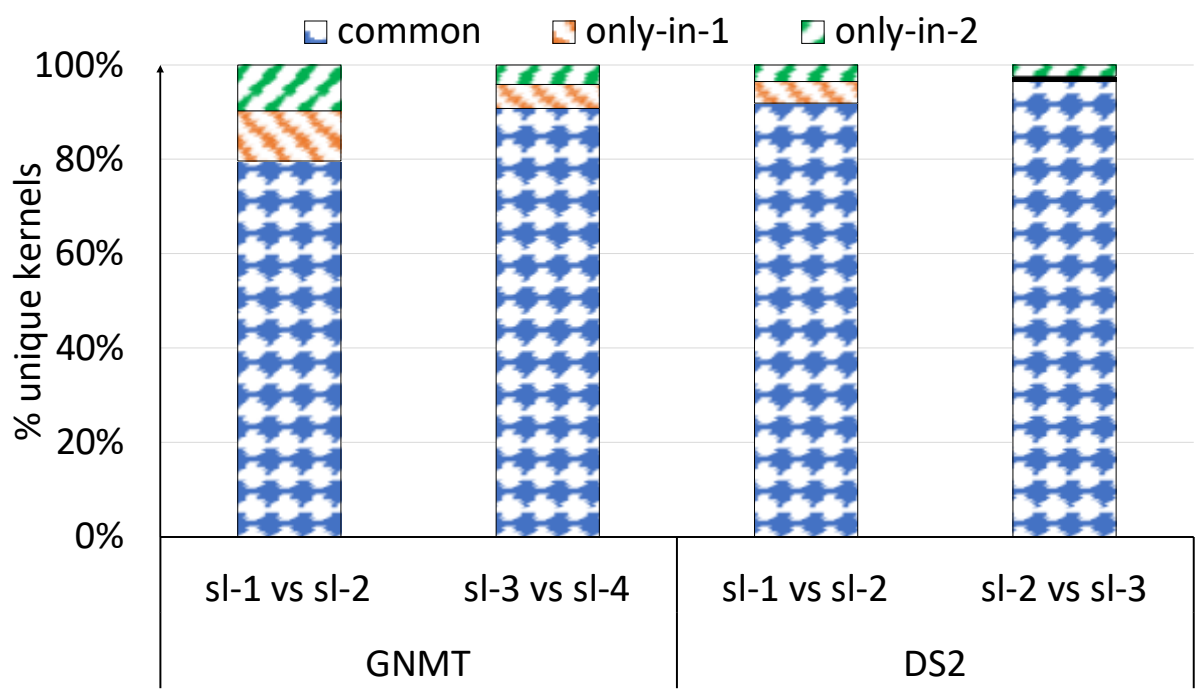

Figure 5: The types of unique kernels differ based on sequence length.

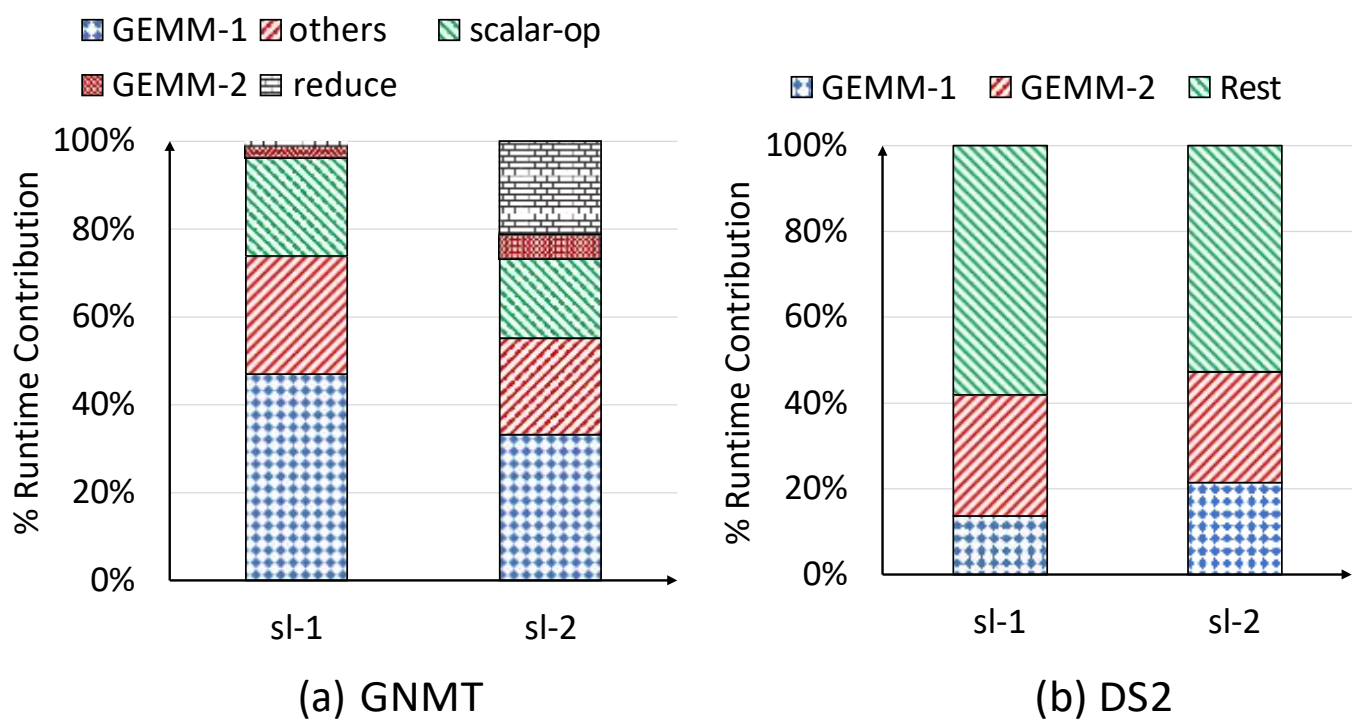

Figure 6: Kernel distribution differs based on sequence length.

Third, and most importantly, in a heterogeneous SQNN, some layers (e.g., attention, convolution, fully-connected) are executed a fixed number of times per iteration and some (e.g., LSTMs, GRUs) are executed in proportion to the sequence length, demonstrating that the proportion of these layers and their respective operations vary across iterations. This in combination with the points above, causes the kernel distribution of operations to differ across iterations as depicted in Fig. 6. The contributions of kernels "GEMM-1" and "reduce" to overall runtime differ significantly based on the iteration's SL in GNMT.

Thus, we make the following key observations of how SL impacts the execution profile of SQNN training iterations:

**Key observation 1:** *Sequence length can differ across iterations and dictates the proportion of operations in an iteration.*

**Key observation 2:** *The total number and type of kernels invoked differ based on the iteration's sequence length.*

**Key observation 3:** *A given kernel can have different input dimensions across iterations and, thus, contribute to the overall execution profile differently.*

*2) Training Dataset:* Multiple datasets are often available to train a given DNN. As discussed in Section II, the dataset dictates the number of iterations within a single training epoch and also the iteration inputs. As such, we observe that representative training iterations are largely tied to the underlying training dataset. However, the training dataset stays constant across all epochs of a single training run. Thus, considering iterations within one epoch is sufficient for identifying a representative set of iterations to profile the entire training. Note that this is independent of whether all epochs execute the iterations in the same order or not.

**Key observation 4:** *Since the dataset is constant during training, considering iterations within a single training epoch is sufficient to generate a representative training phase.*

*3) Iteration Temporal Placement:* As discussed above, the input SL is the key determinant of execution time of an iteration. Thus, in the absence of data-dependent optimizations (e.g., exploiting sparsity, which we do not consider in this work), the behavior of all iterations with a given SL will largely stay the same.

**Key observation 5:** *Unless data-dependent optimizations are used, considering iterations corresponding to unique sequence lengths suffices to generate a representative training phase.*

*4) Vocabulary:* The vocabulary of a dataset in sequence-based networks refers to the unique set of symbols (e.g., words) that appear in a given dataset. The vocabulary size remains fixed across iterations of a training phase and has a considerable effect on the execution time (e.g., lookup time when converting symbols to vectors, input dimensions to operations). Therefore, while sampling training iterations (which may refer to a subset of the dataset), it is important to keep the vocabulary size unchanged to preserve the representativeness of the iterations.

**Key observation 6:** *Since the dataset's vocabulary determines a considerable fraction of per-iteration execution time, we must use the full vocabulary size of the original dataset.*

### C. Non-Training Phase Computations

While DNN training largely comprises training iterations, there are also other computations.

*1) Evaluation Phase:* DNN training includes an evaluation phase at the end of every epoch to determine if a desired level of accuracy has been reached and training can be terminated. The evaluation phase has an independent dataset associated with it and is typically very small compared to the training phase. Unsurprisingly, empirically, we observe that it only takes up to 2-3% of the total training time and, thus, can be ignored when creating a representative execution profile of the training run.

*2) Autotune:* Most high-level MI software frameworks employ an 'autotune' phase at the beginning of a training run

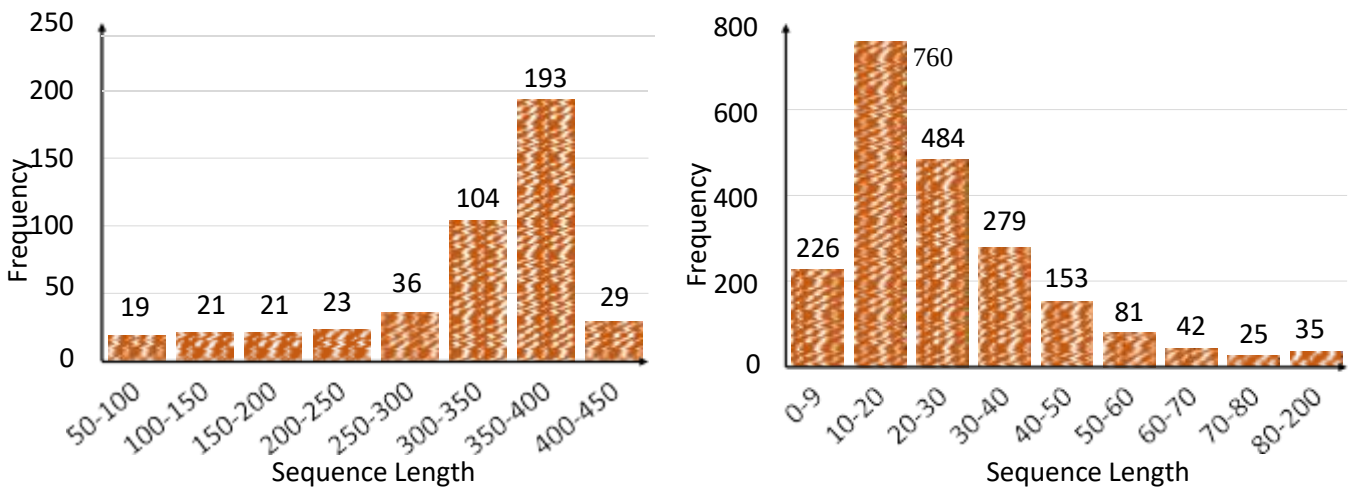

(a) DS2 (b) GNMT

Figure 7: Histogram of SQNN sequence lengths.

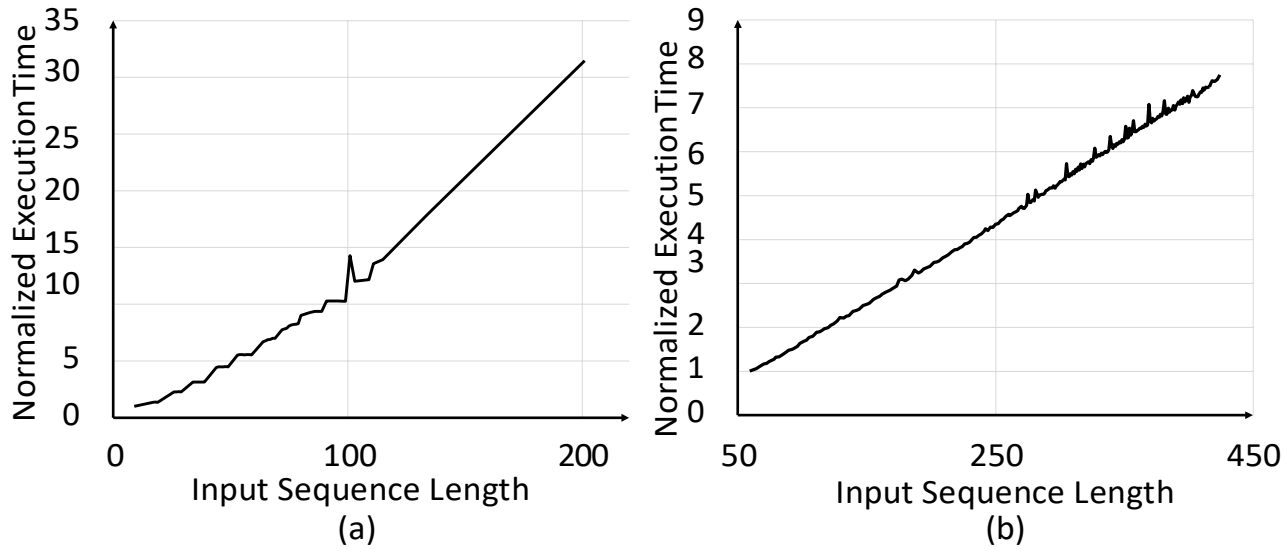

Figure 9: Runtime vs sequence length for (a) GNMT and (b) DS2.

to identify the optimal kernel to run for each computation in the network. It is usually an expensive process and affects the runtime of the first iteration (CNNs) or epoch (SQNNs). However, since autotune only runs once, we can easily ignore it when creating a representative training run.

## V. SeqPoint: Representative Iterations for SQNNs

### A. Challenge: Large Sequence Length Space

In Section IV we analyzed SQNN training and identified several key factors that affect identifying representative iterations. In particular, SL is the key determining factor for variations in execution profile of training iterations. Thus, to select representative iterations of an SQNN training phase, in theory we could include all unique SLs in the training run. However, as Fig. 7 shows, this is challenging because representative datasets for complex SQNNs like DS2 and GNMT have a large number of unique SLs. Consequently, including all unique SLs would lead to a representative set with up to half of all iterations in an epoch (e.g., DS2 with the LibriSpeech [20] 100 hours dataset).

Moreover, the SLs in a given training run are also a function of the batch size. Since most SQNNs pick a single SL (often the maximum SL within the batch of inputs) for an iteration, smaller batch sizes have more unique SLs. Thus, simply selecting all unique SLs is not sufficient, and additional work is needed to identify a smaller set that retains the representativeness of using every unique SL.

### B. SeqPoint Overview

Although SQNNs have a large number of unique SLs (Section V-A), each with a unique execution profile (Section IV-B1), similarly sized SLs have similarity in their execution profiles. Fig. 8 shows that SLs that are close to each other (e.g., 87 and 89, or 192 and 197), have similar kernel distributions. Furthermore, Fig. 9 shows that similarly sized SLs also have similar runtimes. We propose to exploit this similarity to create a smaller, yet still representative set of training iterations, while also taking inspiration from the well-known SimPoint methodology [4].

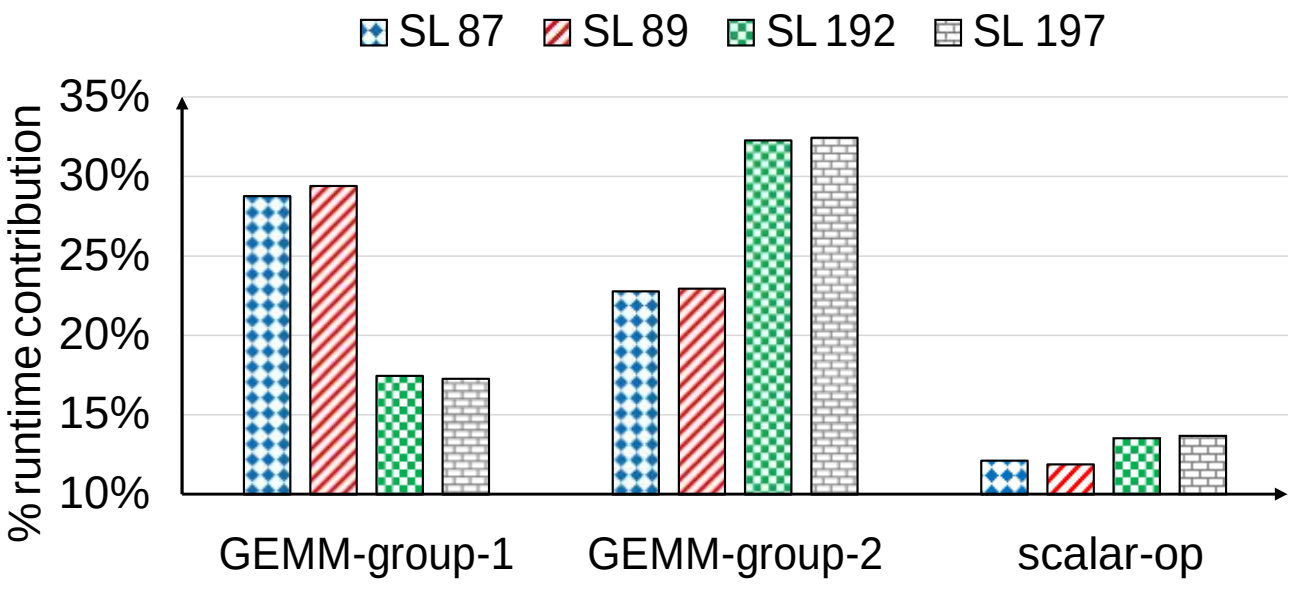

Figure 8: Execution profile with varying sequence length for GNMT.

SimPoint divides program execution into slices and represents each slice with an architecture independent metric: basic-block vector (BBV) which comprises basic blocks and their counts. It then uses clustering over the BBVs and selects a single representative of each cluster termed as SimPoint. In addition, it assigns weights to each SimPoint. Program behavior under SimPoint is then the weighted average of behaviors of individual SimPoints.

In a similar vein, we exploit similarity in SLs to bin them and select a single SL as the representative of each bin, which we term as SeqPoint. In addition, similar to SimPoint, we also assign weights to each of the SeqPoints. The behavior of the entire training run is then a weighted average of all the SeqPoints. Overall, we use a SimPoint-like strategy to create a small, representative subset of the overall training run that is practical to profile and analyze.

### C. SeqPoint Mechanism

Fig. 10 depicts our SeqPoint mechanism. As illustrated in the flowchart, we first execute a single epoch of the SQNN training with the desired network, dataset, and batch size and log all the unique SLs exercised along with the runtime of the respective iterations (1). We also log the training time of the epoch. If desired, to control the training duration, the user can set a threshold, $n$, which decides the number of unique SLs to be included in the representative training run. If the total number of unique SLs is less than this threshold ($n = 10$ for our purposes), we include all unique SLs as SeqPoints.

However, if the number of unique sequence lengths is more than $n$, we bin the observed SLs into $k$ buckets ($k = 5$, initially) each corresponding to a different SL range (2). Our binning of contiguous sequence length ranges is driven by the fact that SLs in close proximity to one another behave similarly (Section V-B).

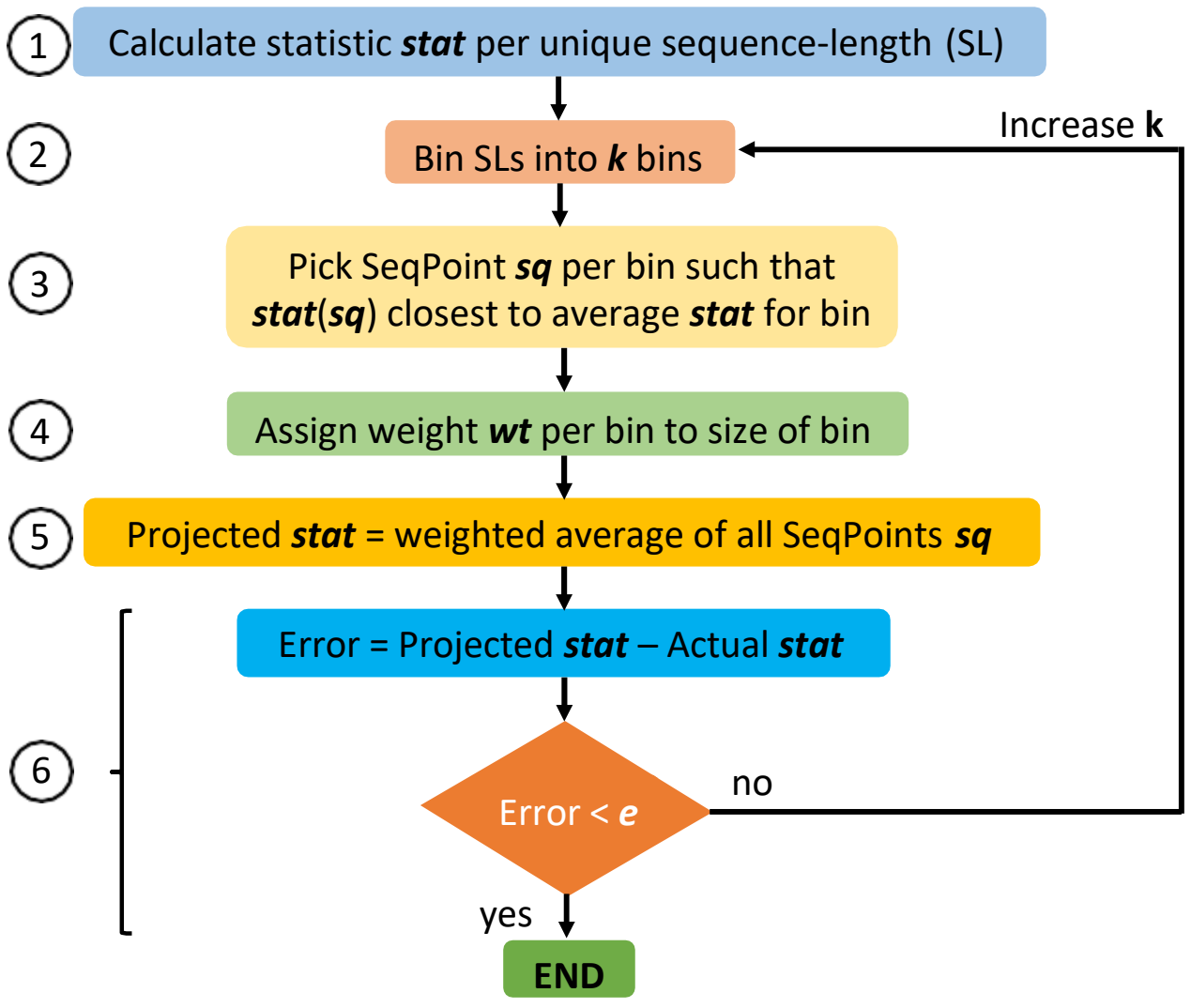

Figure 10: SeqPoint overview.

Next, we pick as the representative from each bin the SL whose runtime ($s$) is closest to the average runtime of the bin and consider it as a SeqPoint (3). This choice exploits the near-linear relationship between runtime (and other statistics) and sequence length within a bin (as shown in Fig. 8 and 9), as well as the fact that iteration runtime is a good enough proxy of the program execution behavior.

In the absence of binning, we assign each SeqPoint a weight ($w$) equal to the frequency of its occurrence in one epoch of training. In the presence of binning, the weight assigned is the size of the bin the SeqPoint belongs to (4). Next, to evaluate the accuracy of the selected SeqPoints, we calculate the weighted sum of the runtimes of each SeqPoint as follows (5):

$$Predicted\ Statistic = w_1 * s_1 + w_2 * s_2 + .. + w_k * s_k \quad (1)$$

If the error between the predicted and actual runtime exceeds an error threshold $e$ (specified by the user, 6), we increment $k$ by one and repeat steps 2 to 6 until the threshold is met. Note that, to predict statistics that are ratios (e.g., throughput, IPC), the value in Equation 1 should be normalized by the sum of all weights.

Overall, given the architectural independence of the SeqPoint methodology, once the SeqPoints for a given combination of model, dataset, and batch size are identified, they can be used to profile the SQNN on any system setup. Further, while we focus on runtime, the methodology can use any other statistic (or collection of statistics) that varies with SL.

# VI. Evaluation

## A. Hardware & Profiling Setup

Our system consists of an AMD Ryzen™ Threadripper [21] CPU and a Radeon™ Vega Frontier Edition GPU [22]. The GPU has 64 compute units (CUs) and 16GB of HBM2 [23]. Our software stack comprises Ten- sorFlow [24] built on top of the AMD ROCm platform [25], and calls into MIOpen [26] and rocBLAS [27], high- performance machine learning libraries from AMD. We use the Radeon Compute Profiler [28], a performance analysis tool, to gather kernel runtimes and other GPU performance counter data.

## B. Networks and Inputs

We study two state-of-the-art SQNNs (reference networks from the MLPerf benchmark suite [5]): Google's Neural Machine Translation (GNMT), which is used for machine translation, and Baidu's DeepSpeech2 (DS2), which is used for speech recognition. GNMT has three main components: (a) an encoder with seven uni-directional and one bi-directional Long Short Term Memory (LSTM) layers, (b) a decoder with eight unidirectional LSTM layers, (c) an attention network, which is a feedforward network connecting the encoder and decoder and (d) a fully-connected layer. DS2 has five bi-directional Gated Recurrent Unit (GRU), two convolutional, one fully-connected, and one batch-normalization layers. We use the IWSLT 2015 [29] and LibriSpeech [20] datasets with a batch size of 64 for GNMT and DS2 respectively.

## C. Methodology

**Hardware configurations:** To show the efficacy of SeqPoint, we evaluate its ability to project both the overall program execution behavior and execution speedups under various hardware configurations for the two SQNNs detailed above. Table II lists the hardware configurations we study. We create five different configurations by varying GPU core frequency (GCLK) and number of active CUs, and by enabling or disabling its L1 and L2 caches. Further, we use total training time as a proxy for program execution behavior and study speedups in terms of increase in training throughput (samples/s). Besides their ability to capture execution profile (or change in execution profile), total training time has been the key metric for benchmarking DNN training, while measuring speedups is key to hardware design, making their accurate projections important.

**SeqPoints**: We generate the SeqPoints and their weights for GNMT and DS2 following the steps detailed in Section V-C. Our methodology identified 15 SeqPoints for GNMT and 8 for DS2, respectively. Note that SeqPoints only need to be identified once, and we do so using config #1. Subsequently, only the SeqPoints are executed on the other configurations. Therefore, representative execution profiles of GNMT and DS2 training can be generated by executing only 15 and 8 iterations, respectively.

**SeqPoint alternatives:** We compare SeqPoint to other alternatives and prior approaches in our evaluation.

*Frequent, Median, Worst:* Prior work [30] used a single iteration as a proxy for the entire training run. By harnessing our insight that SL is a key factor which causes heterogeneous iterations in SQNNs, we devise three strategies to select a single iteration as a representative. `Frequent` selects the most frequently occurring SL, as it has the most likelihood of being picked in a random selection. `Median` selects an iteration with the median SL. Finally, `worst` selects an iteration with the worst-case error to provide a bound on possible error when arbitrarily selecting a single iteration.

Table II: Configurations used to evaluate SeqPoint

| Config | GCLK | #CU | L1 $ | L2 $ |
|---|---|---|---|---|
| **#1** | 1.6 GHz | 64 | 16 KB | 4 MB |
| **#2** | **852 MHz** | 64 | 16 KB | 4 MB |
| **#3** | 1.6 GHz | **16** | 16 KB | 4 MB |
| **#4** | 1.6 GHz | 64 | **0 KB** | 4 MB |
| **#5** | 1.6 GHz | 64 | 16 KB | **0 MB** |

*Prior*: `Prior` uses a sampling-based approach [1] that samples 50 iterations after a fixed warmup period.

### D. Projecting Program Execution Behavior

As discussed in Section VI-C, we use total training time as a proxy for program execution behavior. Fig. 11 and Fig. 12 show the error in projecting the total training time of DS2 and GNMT incurred by SeqPoint (using equation. 1) and its alternatives (by multiplying average iteration time with the number of iterations in an epoch) for the configurations in Table II.

As Fig. 11 and Fig. 12 depict, while we identified SeqPoints using only config #1, they can accurately project training time across a variety of system parameters resulting in geomean errors of 0.11% and 0.53% for DS2 and GNMT, respectively, across the configurations evaluated. This shows that our methodology allows for the SeqPoints to be identified once and be used repeatedly to make accurate program behavior projections across a wide range of architecture and software stack variants.

Moreover, Fig. 11 and Fig. 12 show that SeqPoint alternatives which use a single training iteration to make projections have higher errors. For example, `frequent`, despite being the most frequent SL, has high error (20-35%) and thus is not very representative of the full training run. This is due to the fact that the most frequently occurring SL is not necessarily representative of the distribution of training iterations. Similarly, selecting `median` results in an error as high as 10%.

Despite the projection errors, both `frequent` and `median` were careful selections for a representative iteration based on our understanding of the underlying SL distribution. Selecting an arbitrary, fixed iteration is fraught with higher risk of projection errors as illustrated by `worst` in both the figures.

Fig. 11 and Fig. 12 show that `prior` results in lower errors (about 6%) for DS2 for certain configurations, but performs poorly both for other configurations and GNMT in general. `Prior`'s low error for certain configurations is a consequence of an artifact of DS2's computation: DS2 sorts SLs in the first training epoch, leading to `prior` selecting a set of iterations whose runtimes dominate the training run. Nevertheless, SeqPoint outperforms `prior` by over 5% and, more importantly, does so while running one-third and one-sixth of the iterations as compared to `prior` for GNMT and DS2, respectively.

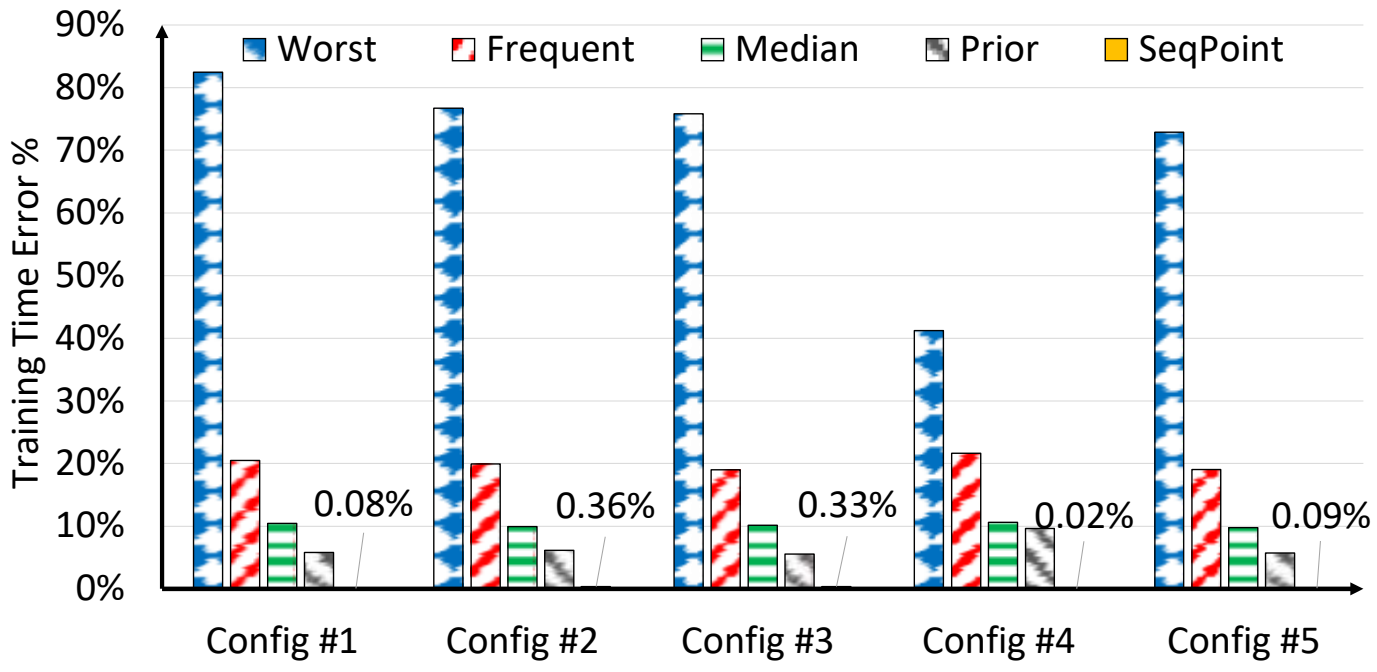

Figure 11: Error in total training time projections for DS2.

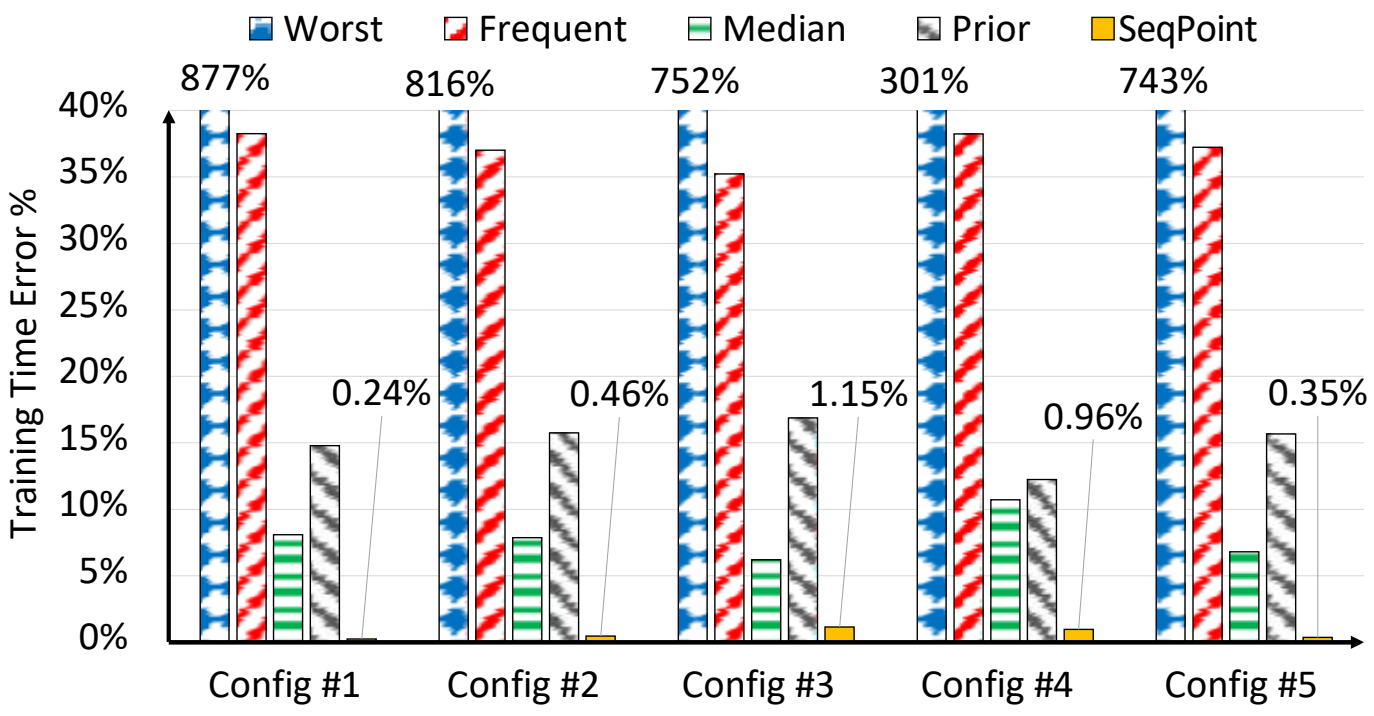

Figure 12: Error in total training time projections for GNMT.

### E. Projecting Performance Speedups

Measuring speedups with change in architectural features is key to hardware design. Fig. 13 and 14 however, show that the speedups of training iterations can vary significantly (by up to 45% for DS2 and 30% for GNMT) due to the variations in sensitivity of different sequence length iterations to the features being changed. Note that this sensitivity can be higher for other hardware features excluded from this study due to the limitations of real hardware analysis. These figures further emphasize the need to pick representative training iterations of an SQNN for evaluating hardware design changes. Therefore, we next evaluate SeqPoint's ability to project speedups as we vary hardware configurations. To do so, we plot the error (delta) in projecting percentage throughput (samples/s) change of end-to-end training phase of the networks between config #1 and other configurations under study.

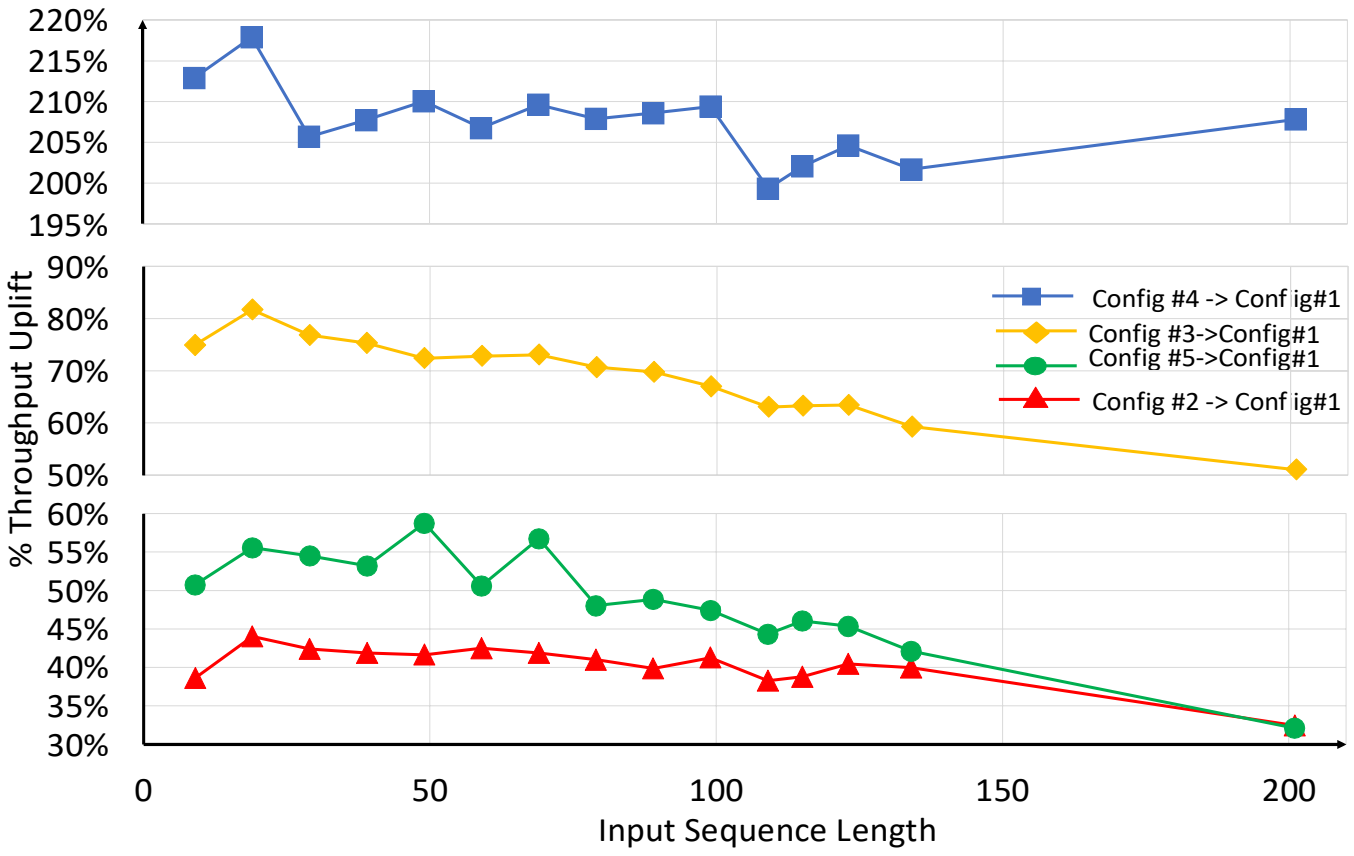

Figure 13: Sensitivity to GCLK, CU count, L1 cache and L2 cache of different sequence length iterations in GNMT.

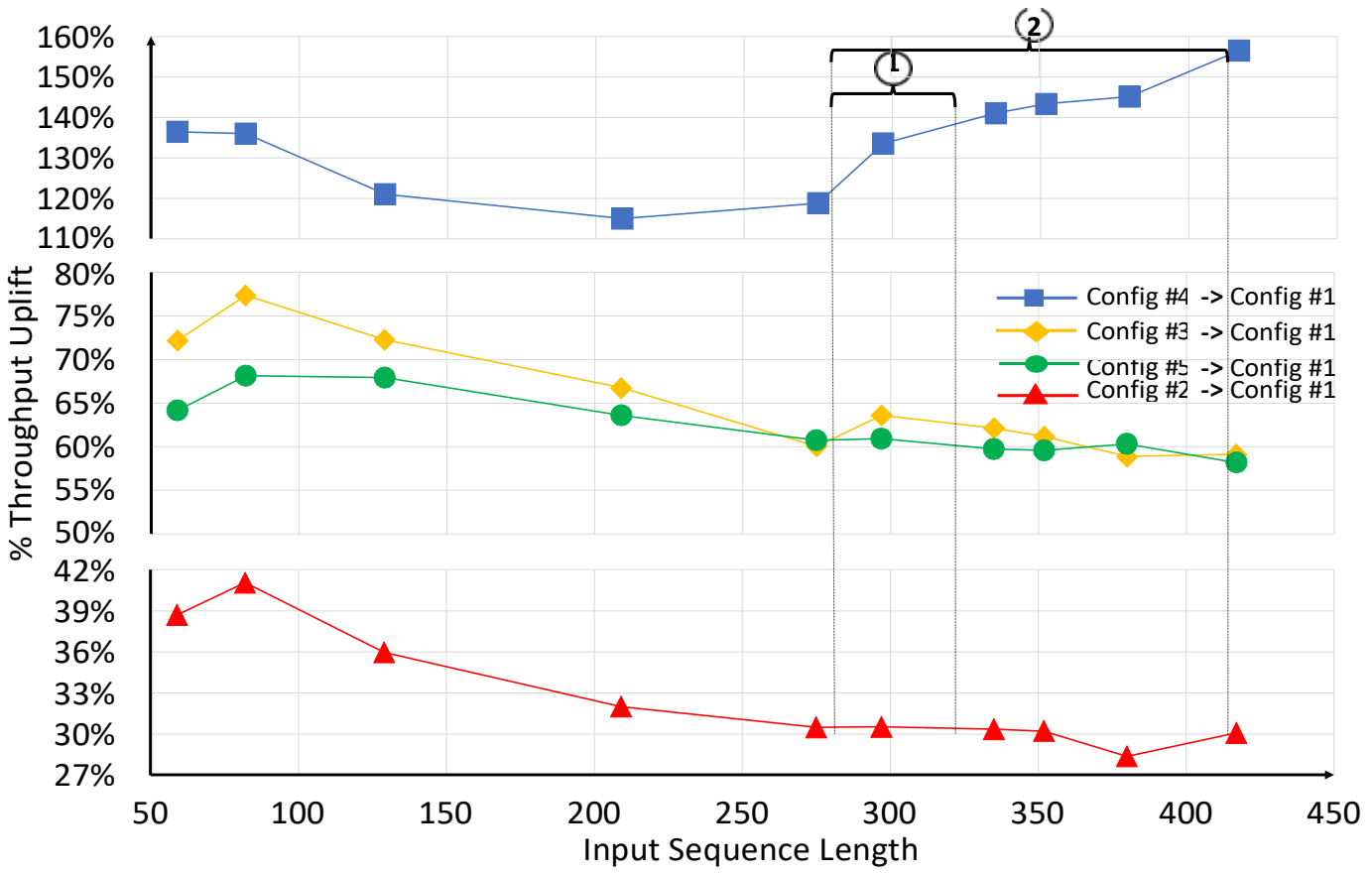

Figure 14: Sensitivity to GCLK, CU count, L1 cache and L2 cache of different sequence length iterations in DS2.

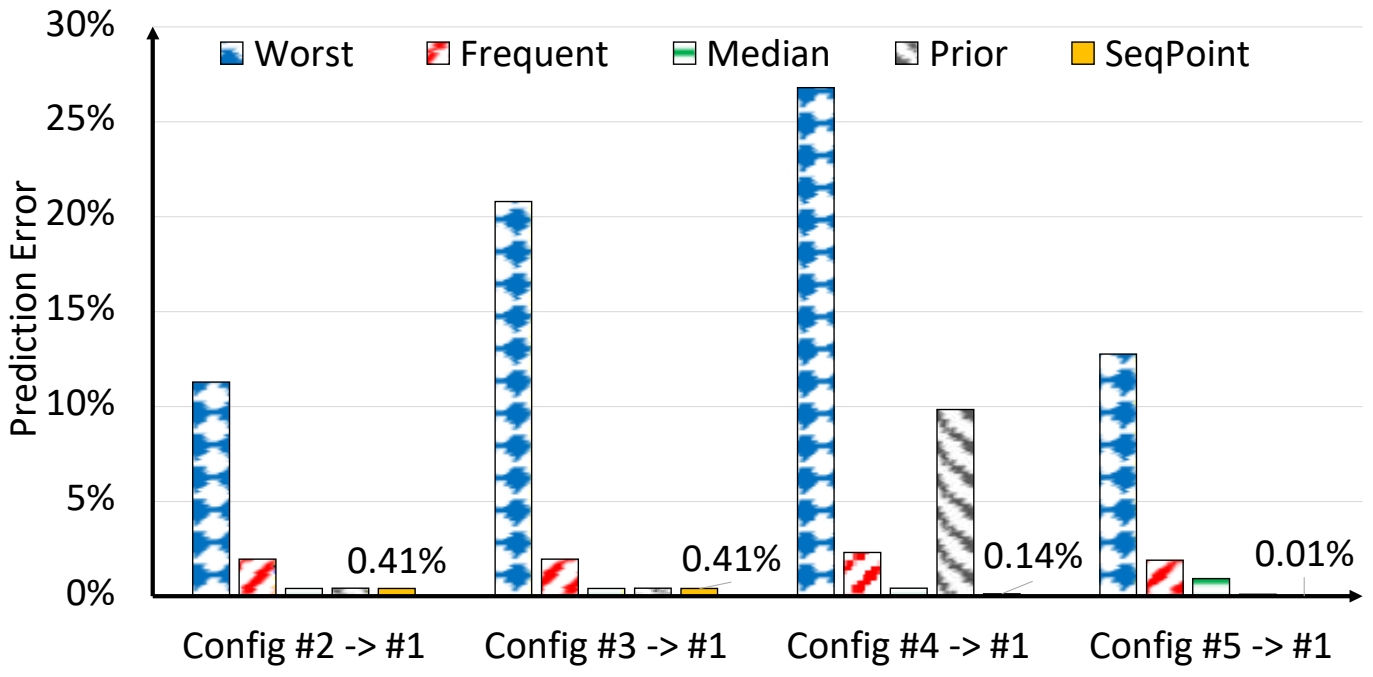

Figure 15: Error in performance speedup projections for DS2.

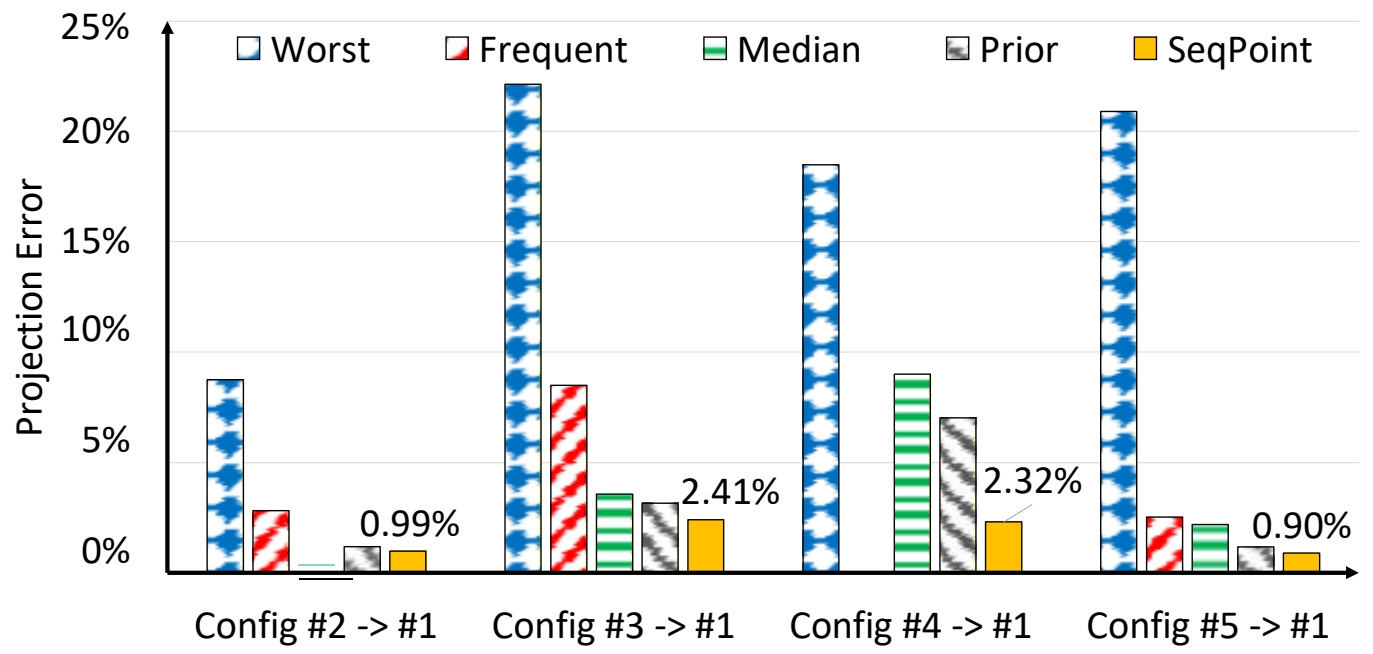

Figure 16: Error in performance speedup projections for GNMT.

Fig. 15 and Fig. 16 show that SeqPoint outperforms all studied alternatives in projecting speedups with geomean errors of 0.13% and 1.50% for DS2 and GNMT, respectively. This further demonstrates SeqPoint methodology's ability to be representative of the entire training run.

Among the SeqPoint alternatives that select single iterations, we observe that while `median` and `frequent` perform worse than SeqPoint, their errors are sometimes within acceptable margins (e.g., 2.5% for DS2). This is due to the fact that both `median` and `frequent` select SLs which are exercised often. Combined with the SL distribution skew in DS2 (Fig. 7), this enables them to accurately predict the relative variation across configurations reflected in the speedups. With a more uniform SL distribution, as is the case for GNMT, `median` and `frequent` exhibit higher errors of up to 9%. Further, as in Section VI-D, `worst` shows the perils of selecting an arbitrary training iteration: errors as high as 22% and 27% for GNMT and DS2 respectively.

We observe in Fig. 15 and Fig. 16 that `prior` does as well as SeqPoint in all cases except when predicting config #4 to #1 speedup for DS2. The region `prior` picks its iterations from is depicted by ① in Fig. 14. The figure also shows that region ② of which ① is a subset, has a constant (and given the skew in Fig. 7(a), also close to overall) uplift for all configs *but* config #4, thus, leading to higher errors for `prior` in projecting config #4 to #1 uplift. This shows that the errors with `prior` can be higher if variations in speedups across sequence lengths is larger.

Furthermore, given `prior`'s design choice of simply picking a subset of the SL space (set of contiguous iterations), its manifested error can be higher when the sequence lengths present in this contiguous chunk are not diverse (due to the non-deterministic order in which models execute different sequence length inputs). This can further lower the representativeness of `prior`.

This further underscores the need to carefully select iterations from the entire SL space of the dataset as our proposed SeqPoint methodology does. Moreover, SeqPoint requires significantly fewer iterations compared to `prior` (up to 6× fewer for DS2) to achieve the above accuracy.

## F. Profiling Speedups

A key benefit of SeqPoint is that it reduces the time required to profile an end-to-end SQNN model training from hours/days to mere minutes/seconds while being extremely accurate. By carefully selecting representative iterations, SeqPoint reduces profiling overheads by 40x and 72x, for GNMT and DS2 respectively. Moreover, given each SeqPoint is an independent iteration, they can be executed in parallel (on different machines) which further speeds up profiling by 214x and 345x, for GNMT and DS2 respectively.

Finally, while we have evaluated SeqPoint only for smaller datasets (LibriSpeech's 100 hours dataset [20] and IWSLT15 dataset [29] for DeepSpeech2 and GNMT, respectively), applying SeqPoint to larger datasets such as the LibriSpeech 500 hours and WMT16 [31], which we observed to have similar SL ranges to the evaluated shorter datasets (data omitted for brevity), can lead to much higher speedups than what we observe for these smaller ones.

## VII. Discussion

### A. Enabling Network-level Simulation for SQNNs

Simulating an entire GPU application on a cycle-level simulator [32], [33] is often impractical and this is even more true for long-running SQNN training applications. To aid in successful simulation of long-running applications, prior works have attempted to identify representative regions within applications and porting them to simulators for CPUs [4], [18], [34]–[36] and GPUs [37], [38].

Such techniques, however, require some form of program analysis, simulation, or profiling which can have significant overhead of up to 10 to 30, making them infeasible for SQNN training (which run for hours to days on native hardware). In contrast, our SeqPoint technique brings down SQNN training time to few seconds to minutes, paving a way for such prior techniques to now profile SQNN training and identify representative portions to simulate. Thus, we believe SeqPoint is a stepping stone to enabling network-level simulations of SQNNs

### B. Other SQNNs

While our analysis focuses on two SQNNs, SeqPoint applies to other networks as well. An insight of this work is to identify input SL as a key factor which determines the variations in execution profile (kernel distribution) for training iterations. As such, any SQNN consisting of layers whose computation varies with input SL can benefit from SeqPoint methodology to reduce the representative training runtime. A wide swath of networks fall into this category which employ layers including, but not limited to, attention (e.g., Transformer [39], BERT [40], and GNMT [3]), convolution (e.g., ConvS2S [41], DeepSpeech2 [2]), and RNN, GRU, LSTM (e.g., Seq2Seq [42], ByteNet [43]).

### C. Sophisticated Clustering of SQNN Iterations

We also considered a more sophisticated approach to tame the SQNN training SL space via k-means clustering [44]. In this approach, we applied k-means clustering to execution profiles of all iterations. However, we observed that our simple methodology to bin SLs (Section V-C) performs as well as k-means clustering and, as such, we use the simpler approach. We believe this to be a consequence of the fact that iteration runtime (which we use) is a good proxy for execution profile of SQNN iterations.

### D. Architecture and Software Independence

The SeqPoint methodology we propose relies entirely on the characteristics of the SQNN model and the dataset it is trained on. Therefore, while our system setup consists of AMD hardware/software stack and the TensorFlow framework, the insights we highlight and the methodology we adopt can apply to any other system (e.g., NVIDIA) and/or framework (e.g., PyTorch, Caffe, or MxNet). Further, while we demonstrate the efficacy of SeqPoints in the context of GPUs, since SeqPoint uses architecture independent metrics (e.g., SL), our methodology could also be applied to CPUs and other accelerators.

### E. SQNN Inference

While the focus of our work has been on SQNN training, insights in this work can be useful in the context of SQNN inference as well. Our observation that SL is a key factor that dictates variations between SQNN iterations is equally applicable to inference. Further, our methodology to bin SLs to tame the SL space can also help characterize inference runs in order to optimize for them.

## VIII. Related Work

The growing importance of SQNNs necessitates tractable profiling methodologies for them. Our work primarily addresses this need by carefully selecting representative points in SQNN training to considerably reduce training iterations (by up to two orders of magnitude) needed to faithfully summarize the entire training run.

As discussed in Section III, prior works [1], [30], [45] assume homogeneity in training iterations which works for CNNs, however, as we show is less effective for SQNNs. By being cognizant of heterogeneity in training iterations of SQNNs, our proposed methodology generates a short set of representative training iterations (SeqPoints) that have lower error as compared to these prior techniques.

Other works, side-step end-to-end profiling of SQNN training and instead focus on analyzing individual layers [15], [19] using microbenchmarks such as DeepBench [46]. However, real-world networks such as DS2 and GNMT comprise several different types of layers (e.g., convolution, attention), interactions among which remain uncaptured by these prior techniques. In contrast, by considering entire iterations, SeqPoint captures these interactions.

Finally, as discussed in Section VII-A, although prior works [4], [18], [34]–[38] can identify representative portions in applications with high accuracy, they are infeasible for long training runs. However, SeqPoint, by providing a considerably shorter (but representative) set of training iterations, paves the way for such techniques to be used in architectural simulation of SQNN training.

## IX. Conclusion

Profiling and characterization of SQNN training runs remain challenging given their hours-to-days native runs. In this work, we observe that prior works which characterize SQNNs are oblivious to the heterogeneity in training iterations and, as such, are ill-equipped to create small, representative training runs that faithfully summarize entire training phases. To address this, we first observe that input SL is a key factor that dictates the heterogeneity of SQNN training iterations. Then, we design a new scheme, SeqPoint, that clusters unique SLs and selects representative points in each cluster. We show our identified SeqPoints are representative of the entire training run with low error. Finally, SeqPoint reduces training iterations by up to two orders of magnitude for state-of-the-art, end-to-end SQNNs. Overall, we not only make profiling and characterization of SQNN training tractable but also pave the way for network-level simulations for SQNNs.

## Acknowledgement

The authors thank the anonymous ISPASS reviewers for helping improve this paper. AMD, AMD Ryzen, AMD Radeon, and combinations thereof are trademarks of Advanced Micro Devices, Inc. Other product names used in this publication are for identification purposes only and may be trademarks of their respective companies.